EUROPEAN ORGANIZATION FOR NUCLEAR RESEARCH

# Experimental investigation of transverse spin asymmetries in $\mu$-p SIDIS processes: Sivers asymmetries

The COMPASS Collaboration

**Abstract**

The COMPASS Collaboration at CERN has measured the transverse spin azimuthal asymmetry of charged hadrons produced in semi-inclusive deep inelastic scattering using a 160 GeV $\mu^+$ beam and a transversely polarised $NH_3$ target. The Sivers asymmetry of the proton has been extracted in the Bjorken $x$ range $0.003 < x < 0.7$. The new measurements have small statistical and systematic uncertainties of a few percent and confirm with considerably better accuracy the previous COMPASS measurement. The Sivers asymmetry is found to be compatible with zero for negative hadrons and positive for positive hadrons, a clear indication of a spin-orbit coupling of quarks in a transversely polarised proton. As compared to measurements at lower energy, a smaller Sivers asymmetry for positive hadrons is found in the region $x > 0.03$. The asymmetry is different from zero and positive also in the low $x$ region, where sea–quarks dominate. The kinematic dependence of the asymmetry has also been investigated and results are given for various intervals of hadron and virtual photon fractional energy. In contrast to the case of the Collins asymmetry, the results on the Sivers asymmetry suggest a strong dependence on the four-momentum transfer to the nucleon, in agreement with the most recent calculations.

**The COMPASS Collaboration**

C. Adolph[8], M.G. Alekseev[24], V.Yu. Alexakhin[7], Yu. Alexandrov[15,*], G.D. Alexeev[7], A. Amoroso[27], A.A. Antonov[7], A. Austregesilo[10,17], B. Badełek[30], F. Balestra[27], J. Barth[4], G. Baum[1], Y. Bedfer[22], J. Bernhard[13], R. Bertini[27], M. Bettinelli[16], K. Bicker[10,17], J. Bieling[4], R. Birsa[24], J. Bisplinghoff[3], P. Bordalo[12,a], F. Bradamante[25], C. Braun[8], A. Bravar[24], A. Bressan[25], E. Burtin[22], L. Capozza[22], M. Chiosso[27], S.U. Chung[17], A. Cicuttin[26], M.L. Crespo[26], S. Dalla Torre[24], S. Das[6], S.S. Dasgupta[6], S. Dasgupta[6], O.Yu. Denisov[28], L. Dhara[6], S.V. Donskov[21], N. Doshita[32], V. Duic[25], W. Dünnweber[16], M. Dziewiecki[31], A. Efremov[7], C. Elia[25], P.D. Eversheim[3], W. Eyrich[8], M. Faessler[16], A. Ferrero[22], A. Filin[21], M. Finger[19], M. Finger jr.[7], H. Fischer[9], C. Franco[12], N. du Fresne von Hohenesche[13,10], J.M. Friedrich[17], V. Frolov[10], R. Garfagnini[27], F. Gautheron[2], O.P. Gavrichtchouk[7], S. Gerassimov[15,17], R. Geyer[16], M. Giorgi[25], I. Gnesi[27], B. Gobbo[24], S. Goertz[4], S. Grabmüller[17], A. Grasso[27], B. Grube[17], R. Gushterski[7], A. Guskov[7], T. Guthörl[9], F. Haas[17], D. von Harrach[13], F.H. Heinsius[9], F. Herrmann[9], C. Heß[2], F. Hinterberger[3], N. Horikawa[18,b], Ch. Höppner[17], N. d'Hose[22], S. Ishimoto[32,c], O. Ivanov[7], Yu. Ivanshin[7], T. Iwata[32], R. Jahn[3], V. Jary[20], P. Jasinski[13], R. Joosten[3], E. Kabuß[13], D. Kang[13], B. Ketzer[17], G.V. Khaustov[21], Yu.A. Khokhlov[21], Yu. Kisselev[2], F. Klein[4], K. Klimaszewski[30], S. Koblitz[13], J.H. Koivuniemi[2], V.N. Kolosov[21], K. Kondo[32], K. Königsmann[9], I. Konorov[15,17], V.F. Konstantinov[21], A. Korzenev[22,d], A.M. Kotzinian[27], O. Kouznetsov[7,22], M. Krämer[17], Z.V. Kroumchtein[7], F. Kunne[22], K. Kurek[30], L. Lauser[9], A.A. Lednev[21], A. Lehmann[8], S. Levorato[25], J. Lichtenstadt[23], T. Liska[20], A. Maggiora[28], A. Magnon[22], N. Makke[22,25], G.K. Mallot[10], A. Mann[17], C. Marchand[22], A. Martin[25], J. Marzec[31], T. Matsuda[14], G. Meshcheryakov[7], W. Meyer[2], T. Michigami[32], Yu.V. Mikhailov[21], M.A. Moinester[23], A. Morreale[22], A. Mutter[9,13], A. Nagaytsev[7], T. Nagel[17], T. Negrini[9], F. Nerling[9], S. Neubert[17], D. Neyret[22], V.I. Nikolaenko[21], W.D. Nowak[9], A.S. Nunes[12], A.G. Olshevsky[7], M. Ostrick[13], A. Padee[31], R. Panknin[4], D. Panzieri[29], B. Parsamyan[27], S. Paul[17], E. Perevalova[7], G. Pesaro[25], D.V. Peshekhonov[7], G. Piragino[27], S. Platchkov[22], J. Pochodzalla[13], J. Polak[11,25], V.A. Polyakov[21], J. Pretz[4,e], M. Quaresma[12], C. Quintans[12], J.-F. Rajotte[16], S. Ramos[12,a], V. Rapatsky[7], G. Reicherz[2], A. Richter[8], E. Rocco[10], E. Rondio[30], N.S. Rossiyskaya[7], D.I. Ryabchikov[21], V.D. Samoylenko[21], A. Sandacz[30], M.G. Sapozhnikov[7], S. Sarkar[6], I.A. Savin[7], G. Sbrizzai[25], P. Schiavon[25], C. Schill[9], T. Schlüter[16], K. Schmidt[9], L. Schmitt[17,f], K. Schönning[10], S. Schopferer[9], M. Schott[10], W. Schröder[8], O.Yu. Shevchenko[7], L. Silva[12], L. Sinha[6], A.N. Sissakian[7,*], M. Slunecka[7], G.I. Smirnov[7], S. Sosio[27], F. Sozzi[24], A. Srnka[5], L. Steiger[24], M. Stolarski[12], M. Sulc[11], R. Sulej[30], H. Suzuki[32,b], P. Sznajder[30], S. Takekawa[28], J. Ter Wolbeek[9], S. Tessaro[24], F. Tessarotto[24], L.G. Tkatchev[7], S. Uhl[17], I. Uman[16], M. Vandenbroucke[22], M. Virius[20], N.V. Vlassov[7], L. Wang[2], M. Wilfert[13], R. Windmolders[4], W. Wiślicki[30], H. Wollny[9,22], K. Zaremba[31], M. Zavertyaev[15], E. Zemlyanichkina[7], M. Ziembicki[31], N. Zhuravlev[7] and A. Zvyagin[16]

[1] Universität Bielefeld, Fakultät für Physik, 33501 Bielefeld, Germany[g]
[2] Universität Bochum, Institut für Experimentalphysik, 44780 Bochum, Germany[g]
[3] Universität Bonn, Helmholtz-Institut für Strahlen- und Kernphysik, 53115 Bonn, Germany[g]
[4] Universität Bonn, Physikalisches Institut, 53115 Bonn, Germany[g]
[5] Institute of Scientific Instruments, AS CR, 61264 Brno, Czech Republic[h]
[6] Matrivani Institute of Experimental Research & Education, Calcutta-700 030, India[i]
[7] Joint Institute for Nuclear Research, 141980 Dubna, Moscow region, Russia[j]
[8] Universität Erlangen–Nürnberg, Physikalisches Institut, 91054 Erlangen, Germany[g]
[9] Universität Freiburg, Physikalisches Institut, 79104 Freiburg, Germany[g]
[10] CERN, 1211 Geneva 23, Switzerland
[11] Technical University in Liberec, 46117 Liberec, Czech Republic[h]
[12] LIP, 1000-149 Lisbon, Portugal[k]
[13] Universität Mainz, Institut für Kernphysik, 55099 Mainz, Germany[g]

[14] University of Miyazaki, Miyazaki 889-2192, Japan[l]
[15] Lebedev Physical Institute, 119991 Moscow, Russia
[16] Ludwig-Maximilians-Universität München, Department für Physik, 80799 Munich, Germany[f,l)]
[17] Technische Universität München, Physik Department, 85748 Garching, Germany[f,l)]
[18] Nagoya University, 464 Nagoya, Japan[l]
[19] Charles University in Prague, Faculty of Mathematics and Physics, 18000 Prague, Czech Republic[h]
[20] Czech Technical University in Prague, 16636 Prague, Czech Republic[h]
[21] State Research Center of the Russian Federation, Institute for High Energy Physics, 142281 Protvino, Russia
[22] CEA IRFU/SPhN Saclay, 91191 Gif-sur-Yvette, France
[23] Tel Aviv University, School of Physics and Astronomy, 69978 Tel Aviv, Israel[n]
[24] Trieste Section of INFN, 34127 Trieste, Italy
[25] University of Trieste, Department of Physics and Trieste Section of INFN, 34127 Trieste, Italy
[26] Abdus Salam ICTP and Trieste Section of INFN, 34127 Trieste, Italy
[27] University of Turin, Department of Physics and Torino Section of INFN, 10125 Turin, Italy
[28] Torino Section of INFN, 10125 Turin, Italy
[29] University of Eastern Piedmont, 15100 Alessandria, and Torino Section of INFN, 10125 Turin, Italy
[30] National Centre for Nuclear Research and University of Warsaw, 00-681 Warsaw, Poland[o]
[31] Warsaw University of Technology, Institute of Radioelectronics, 00-665 Warsaw, Poland[o]
[32] Yamagata University, Yamagata, 992-8510 Japan[l]
[a] Also at IST, Universidade Técnica de Lisboa, Lisbon, Portugal
[b] Also at Chubu University, Kasugai, Aichi, 487-8501 Japan[l]
[c] Also at KEK, 1-1 Oho, Tsukuba, Ibaraki, 305-0801 Japan
[d] On leave of absence from JINR Dubna
[e] present address: III. Physikalisches Institut, RWTH Aachen University, 52056 Aachen
[f] Also at GSI mbH, Planckstr. 1, D-64291 Darmstadt, Germany
[g] Supported by the German Bundesministerium für Bildung und Forschung
[h] Suppported by Czech Republic MEYS grants ME492 and LA242
[i] Supported by SAIL (CSR), Govt. of India
[j] Supported by CERN-RFBR grants 08-02-91009
[k] Supported by the Portuguese FCT - Fundação para a Ciência e Tecnologia, COMPETE and QREN, grants CERN/FP/109323/2009, CERN/FP/116376/2010 and CERN/FP/123600/2011
[l] Supported by the MEXT and the JSPS under the Grants No.18002006, No.20540299 and No.18540281; Daiko Foundation and Yamada Foundation
[m] Supported by the DFG cluster of excellence 'Origin and Structure of the Universe' (www.universe-cluster.de)
[n] Supported by the Israel Science Foundation, founded by the Israel Academy of Sciences and Humanities
[o] Supported by Ministry of Science and Higher Education grant 41/N-CERN/2007/0
[*] Deceased

In the late 60's a simple and powerful description was proposed for the nucleon as a stream of partons each carrying a fraction $x$ of the nucleon momentum in a frame where the nucleon momentum is infinitely large. From the dependence of the deep inelastic lepton-nucleon scattering (DIS) cross section on the energy and momentum transfered to the nucleon it was possible to identify charged partons with the earlier postulated quarks, and assess the existence of gluons as carriers of half of the proton momentum.

Since the 90's it is well known that in order to fully specify the quark structure of the nucleon at twist-two level in quantum chromodynamics (QCD) three types of parton distribution functions (PDFs) are required: the momentum distributions $q(x)$ (or $f_1^q(x)$), the helicity distributions $\Delta q(x)$ (or $g_1^q(x)$) and the transversity distributions $\Delta_T q(x)$ (or $h_1^q(x)$), where $x$ is the Bjorken variable. For a given quark flavour $q$, $q(x)$ is the number density, $\Delta q(x)$ is the difference between the number densities of quarks with helicity equal or opposite to that of the nucleon for a nucleon polarised longitudinally, i.e. along its direction of motion, and the transversity distribution $\Delta_T q(x)$ is the corresponding quantity for a transversely polarised nucleon. If the quarks are assumed to be collinear with the parent nucleon, i.e. neglecting the intrinsic quark transverse momentum $\vec{k}_T$, or after integration over $\vec{k}_T$, the three distributions $q(x)$, $\Delta q(x)$ and $\Delta_T q(x)$ exhaust the information on the internal dynamics of the nucleon. On the other hand, from the measured azimuthal asymmetries of hadrons produced in unpolarised semi-inclusive deep inelastic scattering (SIDIS) and Drell-Yan (DY) processes a sizeable transverse momentum of quarks was derived. Taking into account a finite intrinsic transverse momentum $\vec{k}_T$, in total eight transverse momentum dependent (TMD) distribution functions are required to fully describe the nucleon at leading twist [1]. Presently, PDFs that describe non–perturbative properties of hadrons are not yet calculable in QCD from first principles, but they can already be computed in lattice QCD. In the SIDIS cross section they appear convoluted with fragmentation functions (FFs) [2, 3], so that they can be extracted from the data.

A TMD PDF of particular interest is the Sivers function $\Delta_0^T q$ (or $f_{1T}^{\perp q}$), which arises from a correlation between the transverse momentum $\vec{k}_T$ of an unpolarised quark in a transversely polarised nucleon and the nucleon polarisation vector [4]. In SIDIS this $\vec{k}_T$ dependence gives rise to the "Sivers asymmetry" $A_{Siv}$ which is the amplitude of the $\sin\Phi_S$ modulation in the distribution of the produced hadrons. Here the azimuthal angle $\Phi_S$ is defined as $\Phi_S = \phi_h - \phi_s$ with $\phi_h$ and $\phi_s$ respectively the azimuthal angles of hadron transverse momentum and nucleon spin vector, in a reference system in which the z axis is the virtual photon direction and the xz plane is the lepton scattering plane. Neglecting the hadron transverse momentum with respect to the direction of the fragmenting quark, the Sivers asymmetry can be written as

$$A_{Siv} = \frac{\sum_q e_q^2 \cdot \Delta_0^T q \otimes D_q^h}{\sum_q e_q^2 \cdot q \otimes D_q^h} , \tag{1}$$

where $\otimes$ indicates the convolutions over transverse momenta, $e_q$ is the quark charge and $D_q^h$ describes the fragmentation of a quark $q$ into a hadron $h$.

In the very recent years, much attention has been devoted to the Sivers function, which was originally proposed to explain the large single-spin asymmetries observed in hadron-hadron scattering. The Sivers function is T–odd, namely it changes sign under naive time reversal, which is defined as usual time reversal but without interchange of initial and final state. For a long time the Sivers function and the corresponding asymmetry were believed to vanish [5] due to T–invariance arguments. However Brodsky et al. [6] showed by an explicit model calculation that final-state interactions in SIDIS arising from gluon exchange between the struck quark and the nucleon remnant (or initial state in DY) produce a non-zero asymmetry. One of the main theoretical achievements of the recent years was the discovery that the Wilson-line structure of parton distributions, which is necessary to enforce gauge invariance of QCD, provides the possibility for non-zero T–odd transverse momentum dependent (TMD) PDFs. According to factorisation the T–odd PDFs are not universal. The Sivers function can be different from zero but must have opposite sign in SIDIS and DY [7]. A lot of interest in the Sivers function arises also from its relation with orbital motion of quarks inside a transversely polarised nucleon. In particular it was shown [6] that orbital angular momentum must exist if the Sivers function doesn't vanish. Even though no exact relation between Sivers function and orbital angular momentum was derived yet, work is going on, also because the importance of assessing the role of the orbital angular momentum in the nucleon spin sum rule has grown in time (see e.g. [8–11]).

Presently, the measurement of the Sivers asymmetry in SIDIS is the only direct way to assess the Sivers function. It became an important part of the experimental programs of the HERMES and COMPASS experiments, and it will be an important part of future SIDIS experiments at JLab12 [12]. Furthermore, in the near future several experiments using the DY process will address the Sivers function, in particular its sign, in order to establish the prediction of restricted universality [13, 14].

Using a 160 GeV longitudinally polarised $\mu^+$ beam COMPASS measured SIDIS on a transversely polarised deuteron ($^6$LiD) target in 2002, 2003 and 2004. In those data no sizeable Sivers asymmetry was observed within the accuracy of the measurements [15–17], a fact which is understood in terms of a cancellation between the contributions of u- and d-quarks. By scattering the $e^-$ and $e^+$ beams at HERA off a transversely polarised proton target, HERMES measured in 2004 a non-zero Sivers asymmetry for positively charged hadrons [18]. A combined analysis of the COMPASS and HERMES data allowed for a first extraction of the Sivers function for u- and d-quarks [19–21]. Still, as in the case of the Collins asymmetry, measurements on protons at higher beam energies were needed to disentangle possible higher twist effects.

In 2007 COMPASS measured for the first time SIDIS on a transversely polarised proton ($NH_3$) target. The results [22] on the Sivers asymmetry for positive hadrons were found to be different from zero and turned out to be somewhat smaller than the final HERMES data [23]. However the COMPASS results had larger statistical errors and a non-negligible overall scale uncertainty of $\pm 0.01$. A more precise measurement was thus mandatory and the entire 2010 data taking period was dedicated to this purpose.

In this Letter, the results of the 2010 run are presented. They confirm with considerably smaller uncertainties the observation of the 2007 measurements. The higher statistics allow for first studies of the kinematic dependence of the asymmetry in a domain larger than the usual COMPASS DIS phase space.

The COMPASS spectrometer is in operation in the SPS North Area of CERN since 2002. The principle of the measurement and the data analysis were already described in refs. [15–17, 22, 24]. The information on the 2010 run, the amount of data collected, the event reconstruction and selection, the statistics of the final samples, are given in a parallel paper on the Collins asymmetry [25] that was measured using the same data. In order to ensure a DIS regime, only events with photon virtuality $Q^2 > 1$ (GeV/c)$^2$, fractional energy of the virtual photon $0.1 < y < 0.9$, and mass of the hadronic final state system $W > 5$ GeV/c$^2$ are considered. A charged hadron is required to have at least 0.1 GeV/c transverse momentum $p_T^h$ with respect to the virtual–photon direction and a fraction of the available energy $z > 0.2$. This is refered to as "standard sample" in the following.

The Collins and Sivers asymmetries are the amplitudes of 2 of the 8 azimuthal modulations, which are theoretically expected to be present in the SIDIS cross section for a transversely polarised target. They are extracted simultaneously from the same data as explained in ref. [25]. The measured amplitude of the modulation in $\sin\Phi_S$ is $\varepsilon_S = f P_T A_{Siv}$, where $f$ is the dilution factor of the $NH_3$ material, and $P_T$ the magnitude of the proton polarisation. In order to extract $A_{Siv}$, the measured amplitudes $\varepsilon_S$ in each period are divided by $f$ and $P_T$. The dilution factor of the ammonia target is calculated for semi-inclusive reactions [26] and is evaluated in each $x$ bin; it increases with $x$ from 0.14 to 0.17, and it is assumed constant in $z$ and $p_T^h$. The proton target polarisation ($\sim 0.8$) was measured individually for each cell and each period. The results for $A_{Siv}$ from all periods of data taking are found to be statistically compatible and the final asymmetries are obtained by averaging the results from the full available statistics. Extensive studies were performed in order to assess the systematic uncertainties of the measured asymmetries, and it was found that the largest contribution is due to residual acceptance variations within the data taking periods. In order to quantify these effects, various types of false asymmetries are calculated from the final data sample assuming wrong sign polarisation for the target cells. Moreover, the physical asymmetries are extracted splitting the events according to the detection of the scattered muon in the spectrometer (top vs bottom, left vs right). The differences between these physical asymmetries and the false asymmetries are used to quantify the overall systematic point-to-point uncertainties, which are evaluated to be 0.5 times the statistical uncertainties. The only relevant systematic scale uncertainty, which arises from the measurement of the target polarisation, is evaluated to be 3% of the target polarisation.

Figure 1 shows the Sivers asymmetries for positive and negative hadrons extracted from the 2010 proton data as a function of $x$, $z$ and $p_T^h$, where the other two variables are integrated over. For negative hadrons the asymmetry is compatible with zero, while for positive hadrons it is definitely positive and stays positive down to $x \simeq 10^{-3}$, in the region of the quark sea. There is good agreement with the published results from the COMPASS 2007 run [22] but with a considerable reduction of more than a factor of two in the statistical and in the point-to-point systematic uncertainties. Also, the asymmetry for positive hadrons is clearly smaller than the corresponding one measured by HERMES [23]. This fact persists even when considering only events with $x > 0.032$, in the same $x$ range as the HERMES experiment. The asymmetries in this restricted $x$ range are shown as open points in fig. 2.

The correlation between the Collins and the Sivers azimuthal modulations introduced by the non-uniform azimuthal acceptance of the apparatus as well as the correlations between the Sivers asymmetries measured when binning the same data alternatively in $x$, $z$ or $p_T^h$ were already given in ref. [25]. All correlation coefficients are found to be smaller than 0.2 and are relevant only in case of simultaneous fits of the various asymmetries.

In order to further investigate the kinematic dependence of the Sivers asymmetry and to understand the reason of the difference with HERMES, the kinematic domain is enlarged to examine the events with smaller $y$ values (in the interval $0.05 < y < 0.1$), which correspond to smaller $Q^2$ and $W$ values. Additionally, the standard data sample is divided into two parts, corresponding to $0.1 < y < 0.2$ and $0.2 < y < 0.9$. Since at small $y$ there are no low-$x$ data, only events with $x > 0.032$ are used. Figure 3 shows the Sivers asymmetries measured in these three bins of $y$ as a function of $x$, $z$, and $p_T^h$ respectively. No particular trend is observed in the case of the asymmetries for negative hadrons (bottom plots), which stay compatible with zero as for the standard sample. A clear increase of the Sivers asymmetry for positive hadrons is visible for the low-$y$ data. This strong effect can not be due to the slightly different mean values of $x$, since the Sivers asymmetry does not exhibit an $x$ dependence for $x > 0.032$. On the contrary, it could be associated with the smaller values of $Q^2$ and/or with the smaller values of the invariant mass of the hadronic system $W$. A similar dependence of the asymmetries on $y$ was already noticed in the published results from the 2007 data. As can be seen from fig. 4 (left panel), there is a strong correlation between the $y$ and $W$ mean values: the mean values of $W$ in the high $x$ bins are about 3 GeV/c$^2$ for the sample $0.05 < y < 0.1$ and larger than 5 GeV/c$^2$ for the standard sample $0.1 < y < 0.9$ (middle panel of fig. 4). On the other hand, as can be seen in the right panel of fig. 4, bins at smaller $y$ have smaller values of $\langle Q^2 \rangle$. In particular, in each $x$ bin the $Q^2$ mean value decreases by about a factor of 3 for the sample $0.05 < y < 0.1$ with respect to the standard sample. Although the situation might be different in the target fragmentation region [27], in the current fragmentation region the Sivers asymmetry is not expected to depend on $y$ (or on $W$), while some $Q^2$ dependence should exist due to the $Q^2$ evolution of both the FFs and the TMD PDFs.

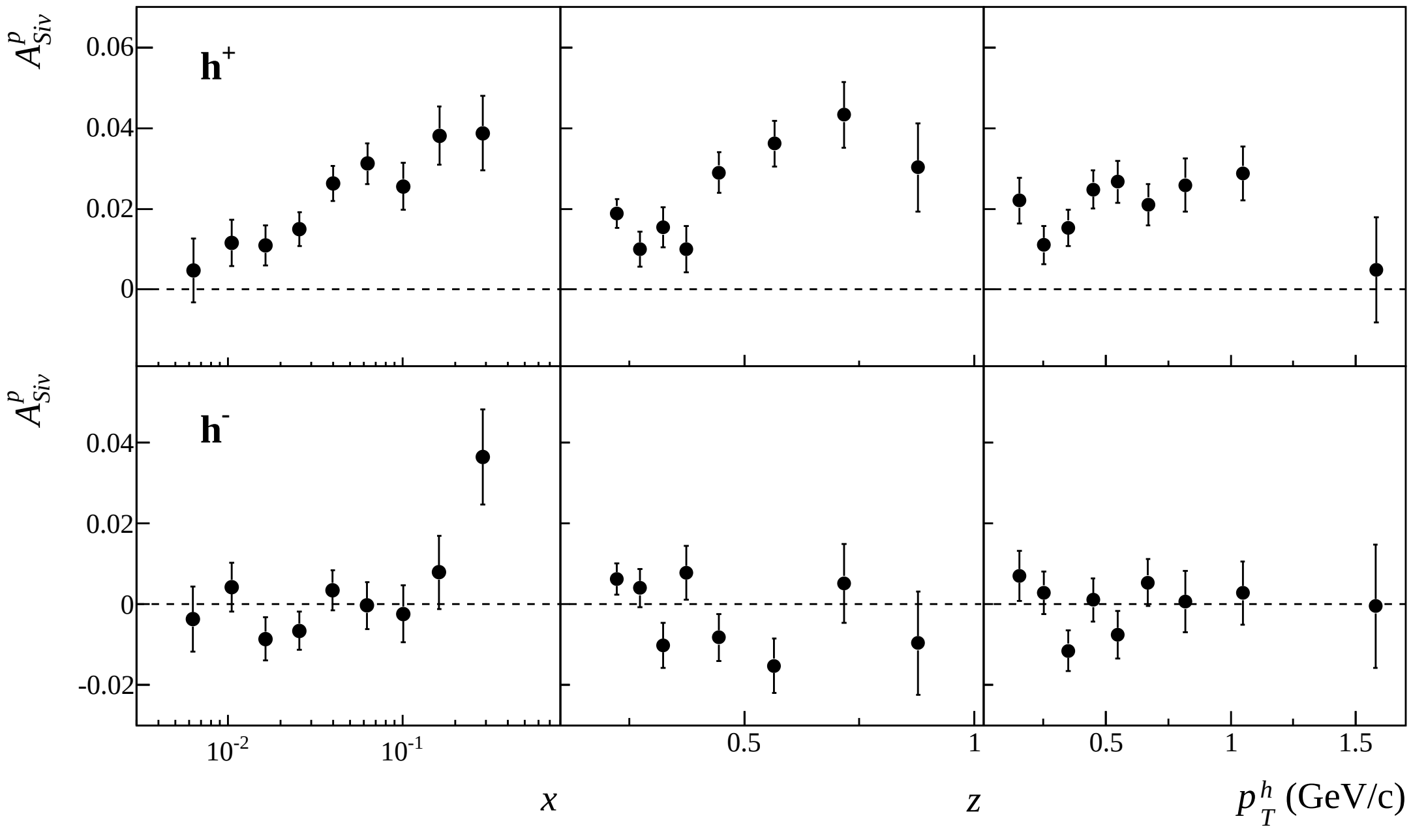

Fig. 1: Sivers asymmetry as a function of $x$, $z$ and $p_T^h$ for positive (top) and negative (bottom) hadrons.

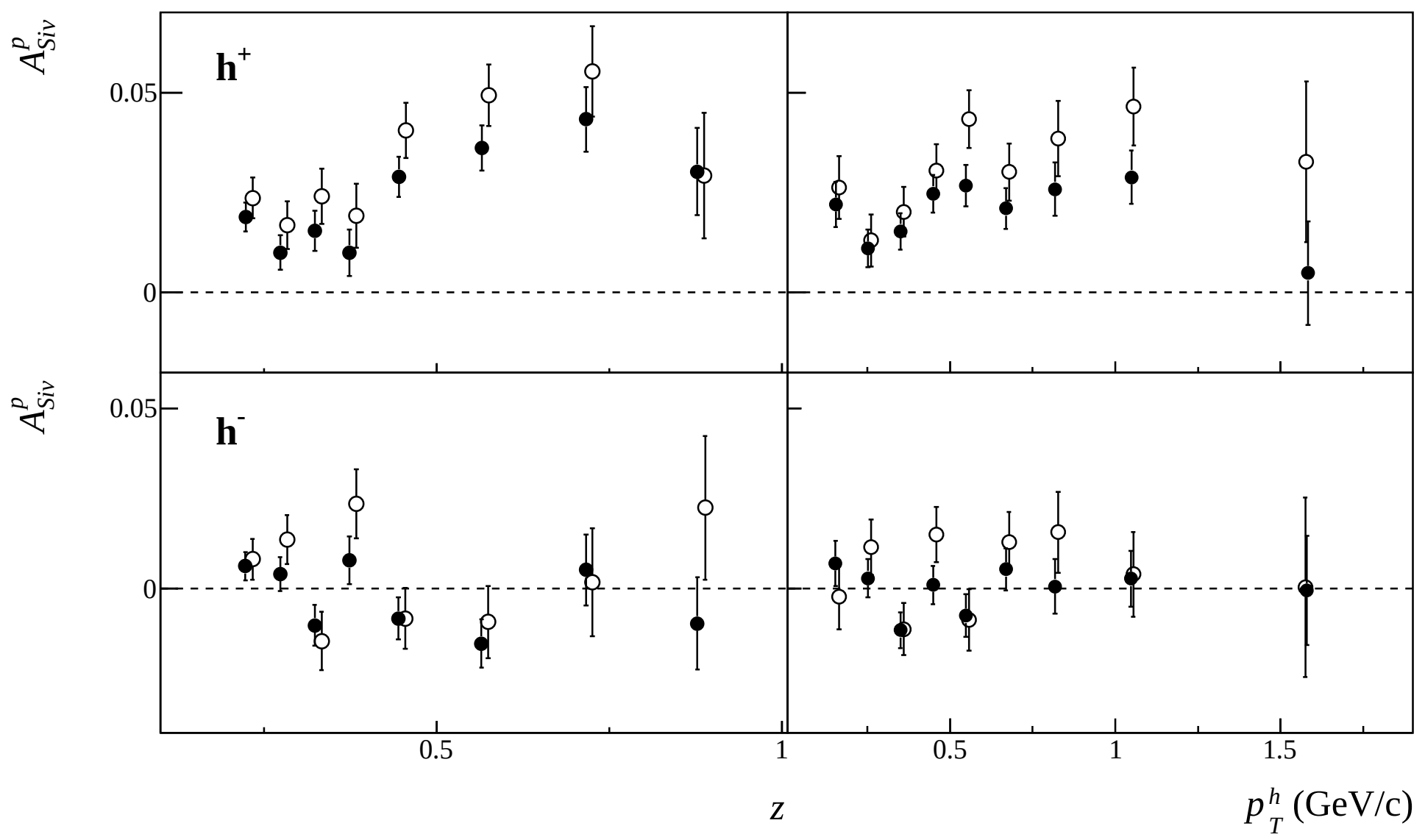

Fig. 2: Sivers asymmetry as a function of $z$ and $p_T^h$ for positive (top) and negative (bottom) hadrons. The open points (○ , slightly shifted horizontally) are the values obtained in the range $0.032 < x < 0.70$. The closed points (●) refer to the full $x$ range and are the same as in fig. 1.

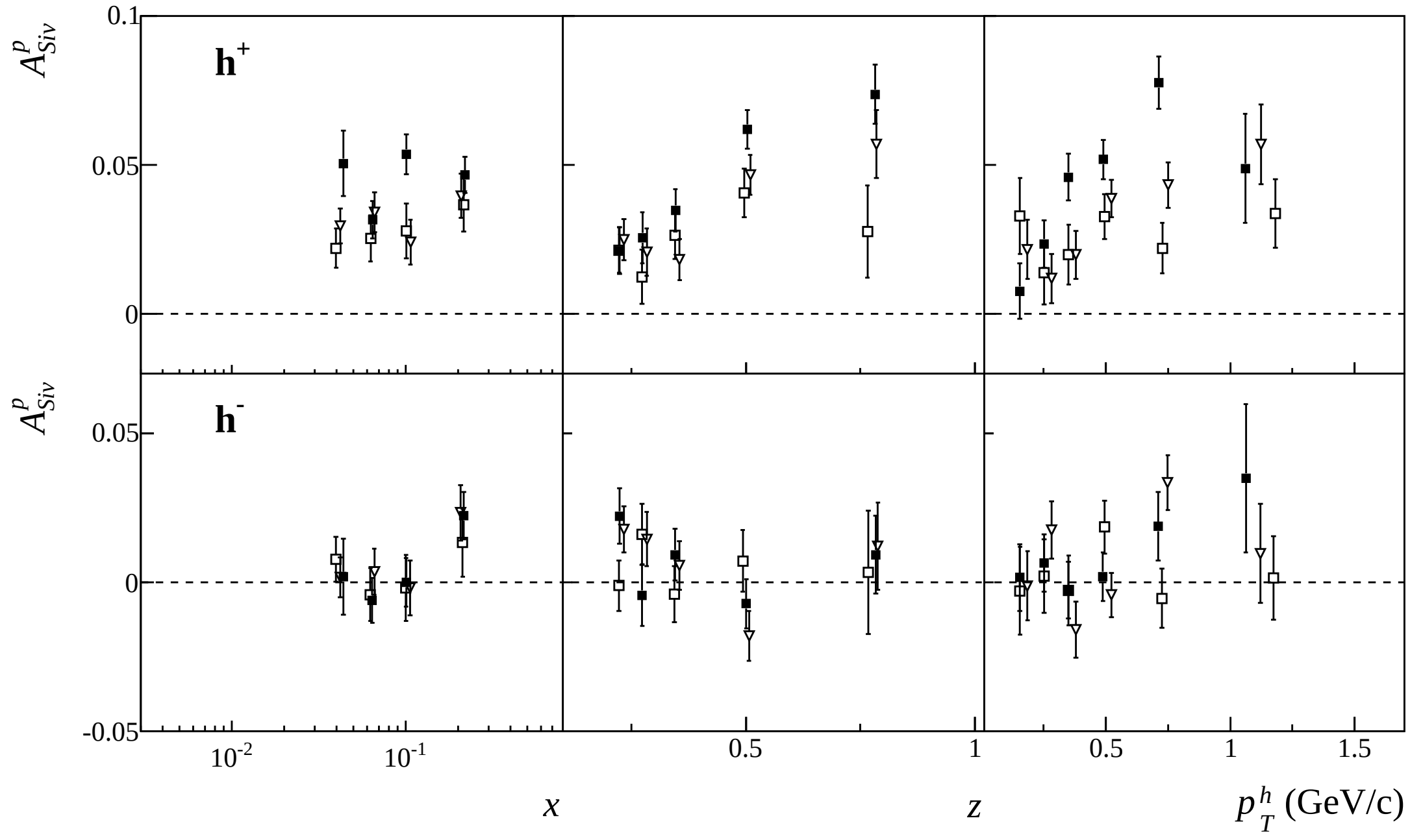

Fig. 3: Sivers asymmetry as a function of $x$, $z$ and $p_T^h$ for positive (top) and negative (bottom) hadrons for $x > 0.032$ in the $y$ bins $0.05 < y < 0.1$ (closed squares, ■), $0.1 < y < 0.2$ (open triangles, ▽, slightly shifted horizontally) and $0.2 < y < 0.9$ (open squares, □) .

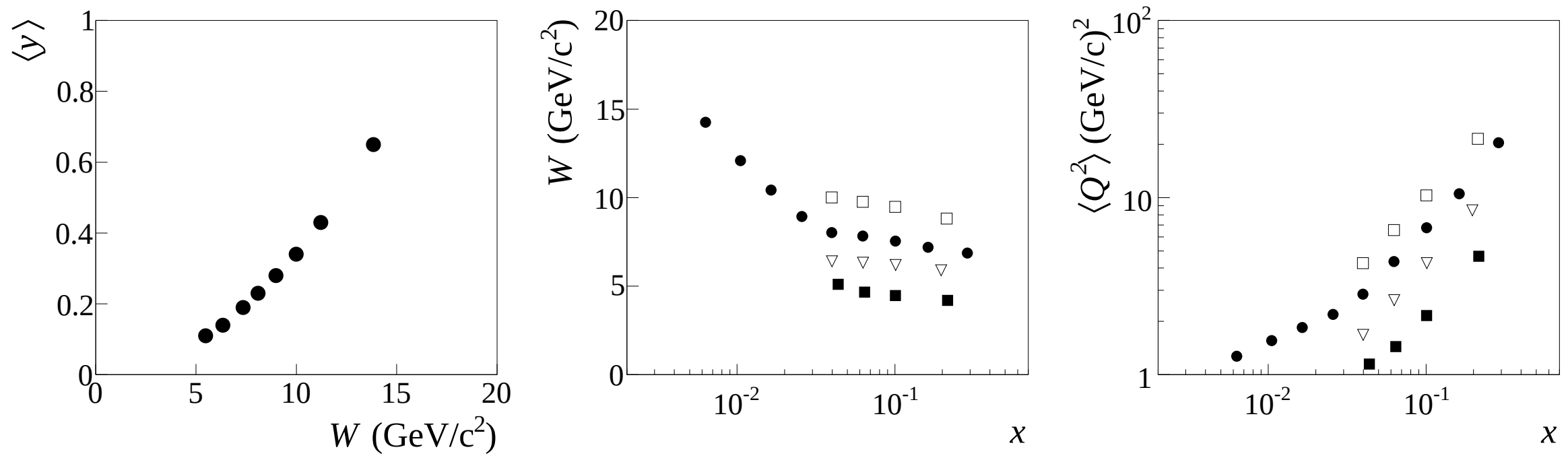

Fig. 4: Left panel: mean value of $y$ vs $W$. Middle panel: mean values of $W$ vs $x$ for the standard sample $0.1 < y < 0.9$ (closed circles, •) and for the samples $0.05 < y < 0.1$ (closed squares, ■), $0.1 < y < 0.2$ (open triangles, ▽), and $0.2 < y < 0.9$ (open squares, □). Right panel: mean values of $Q^2$ vs $x$ for the standard sample $0.1 < y < 0.9$ (closed circles, •) and for the samples $0.05 < y < 0.1$ (closed squares, ■), $0.1 < y < 0.2$ (open triangles, ▽), and $0.2 < y < 0.9$ (open squares, □).

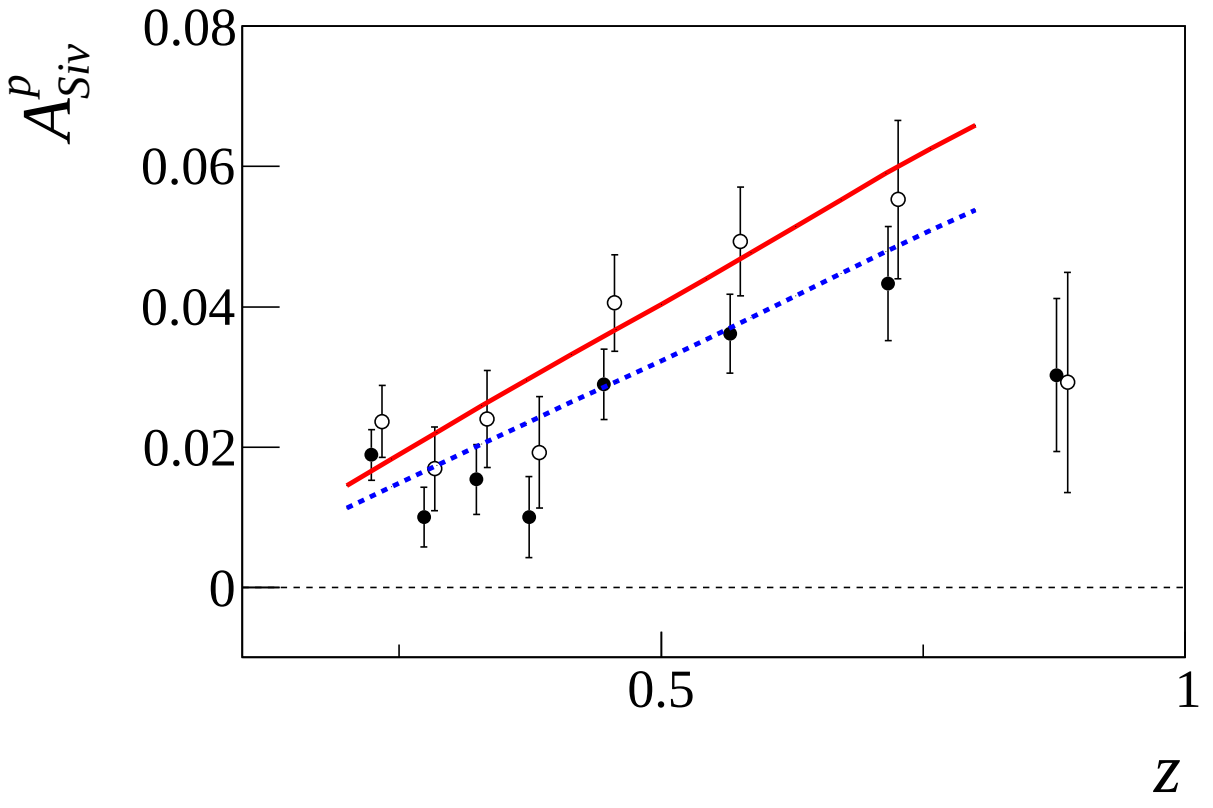

Fig. 5: Comparison between the measured and calculated Sivers asymmetries for positive hadrons as a function of $z$ for $0.1 < y < 0.9$. The closed points (•) refer to the full $x$ range and the open points (○) to the $0.032 < x < 0.70$ range. The curves are from ref. [29].

Very recently first attempts to estimate the impact of the $Q^2$ evolution of the Sivers function [28] led to encouraging results. In ref. [29] the Sivers asymmetry was evaluated for the HERMES kinematic region using the Sivers functions of ref. [30] and then evolved to the COMPASS kinematic region. The measured $z$ dependence of the Sivers asymmetries for $0.1 < y < 0.9$ is compared with the calculated one in fig. 5, for the entire $x$ region and for $x > 0.032$. The linear trend of the data up to $z \simeq 0.75$ is well reproduced, as well as the small increase of the slope for the high $x$ sample. A very recent fit [31] of the HERMES asymmetries [23] and the COMPASS deuteron [17] and proton [32] results given here was performed taking into account the $Q^2$ evolution in all $x$ bins. It reproduces all the data well and provides strong support to the current TMD approach, which foresees a strong $Q^2$–dependence of the Sivers function.

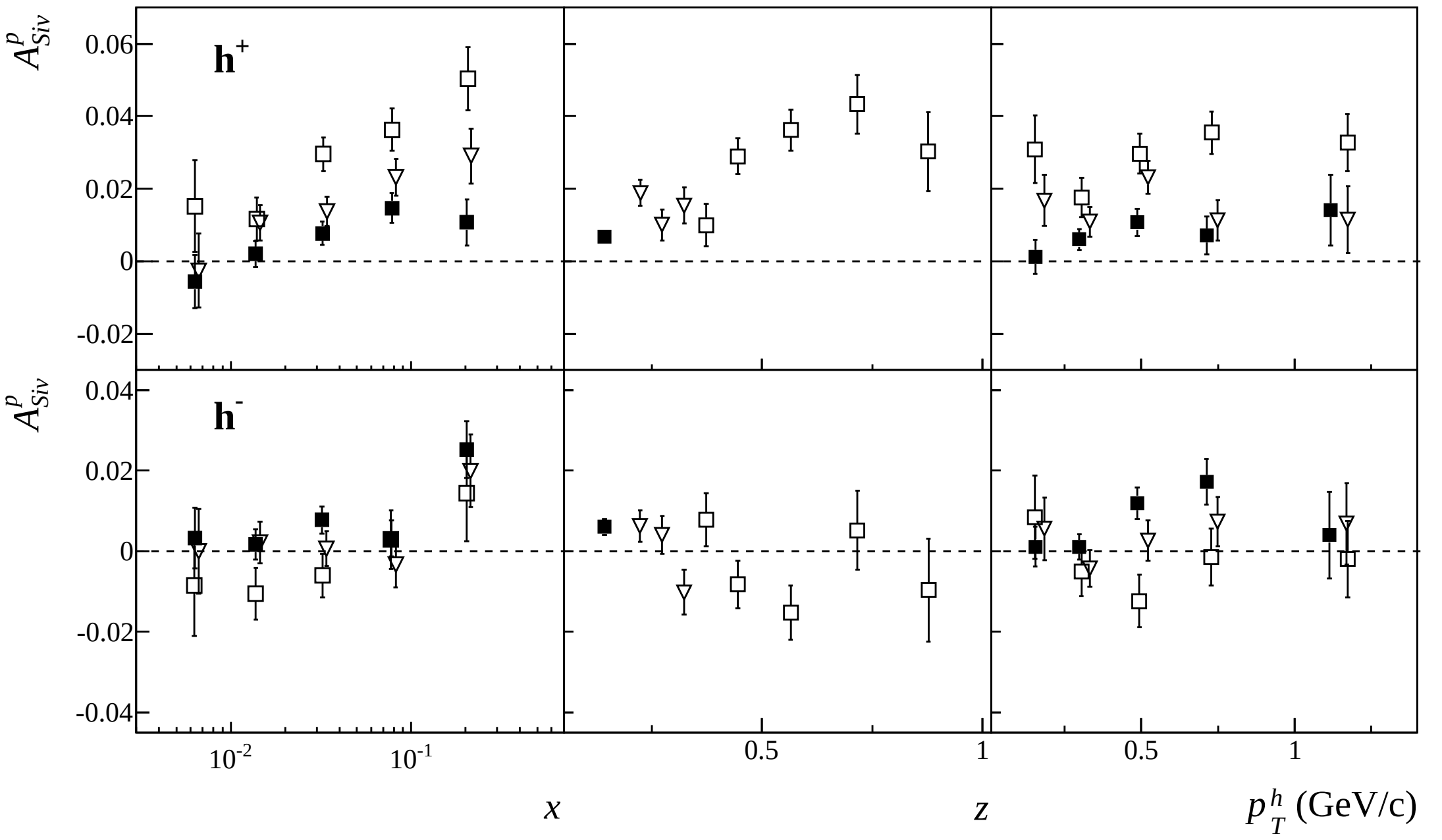

Fig. 6: Sivers asymmetry as a function of $x$, $z$ and $p_T^h$ for positive (top) and negative (bottom) hadrons for $0.032 < x < 0.70$ in 3 different $z$ bins: $0.1 < z < 0.2$ (closed squares, ■), $0.2 < z < 0.35$ (open triangles, ▽, slightly shifted horizontally when plottes vs $x$ and $p_T^h$) and $0.35 < z < 1.0$ (open squares, □).

We have also investigated the behaviour of the Sivers asymmetries at low $z$. Our standard hadron selection requires $z > 0.2$ to stay well separated from the target fragmentation region. In the range $0.1 < z < 0.2$ no effect on $A_{Siv}$ is visible for negative hadrons, but one observes a clear decrease of the asymmetry for positive hadrons. In fig. 6 the data are plotted in 3 different $z$ regions: $0.10 < z < 0.20$, $0.20 < z < 0.35$, and $0.35 < z < 1.00$. While the shape of the asymmetry as a function of $x$ stays the same, the size of the asymmetry shows a clear proportionality with $z$, in qualitative agreement with the expected linear behaviour (see, e.g. [33]).

All the results given in this Letter are available on HEPDATA [34]. The asymmetries for the standard sample as functions of $x$, $z$ and $p_T^h$ have also been combined with the already published results from the 2007 run [22] and are also available on HEPDATA.

In summary, COMPASS has obtained precise results on the Sivers asymmetry in SIDIS using a polarised proton target. A first investigation of its dependence on various kinematic variables shows significant dependences on $z$ and $y$. By now, the Sivers asymmetry for positive hadrons is shown to be different from zero in a broad kinematic range and to exhibit strong kinematic dependences. After two decades of speculations, this is an important new insight into the partonic structure of the nucleon. In the light of the most recent theoretical advances refined combined analyses to evaluate the Sivers function and its dependence on the SIDIS variables are required in order to understand the role of the Sivers function in the various transverse spin phenomena observed in hadron-hadron collisions and in future Drell-Yan measurements.

We acknowledge the support of the CERN management and staff, as well as the skills and efforts of the technicians of the collaborating institutes.